\documentclass[twocolumn,secnumarabic,amssymb, nobibnotes, aps, prd, showpacs]{revtex4}
\usepackage[dvipdf]{graphicx}

\begin{document}

\title{Antiphase Synchronization in Environmentally coupled Rossler Oscillators}

\author{G. Ambika}
\email{g.ambika@iiserpune.ac.in}
\affiliation{Indian Institute of Science Education and Research, Pune-411 021, India} 
\author{Sheekha Verma}
\email{sheekha.iitkgp@gmail.com}
\affiliation{Indian Institute of Technology, Kharagpur-721302, India} 

\begin{abstract}
We study the manifestation of antiphase synchronization in a system of n Rossler Oscillators coupled through a dynamic environment. When the feedback from system to environment is positive (negative) and that from environment to system is negative (positive), the oscillators enter into a state of antiphase synchronization both in periodic and chaotic regimes. Their phases are found to be uniformly distributed over $2\pi$, with a phase lag of $2\pi/n$ between neighbors as is evident from the similarity function and the phase plots. The transition to antiphase synchronization is marked by the crossover of (n-1) zero Lyapunov Exponents to negative values. If the systems are individually in chaotic phase, with strong enough coupling they end up in periodic states which are in antiphase synchronization
\end{abstract}

\pacs{05.45.+b}

\maketitle

%\section{\label{sec:level1}INTRODUCTION}
Synchronization of mutually coupled chaotic systems has been an area of intense research activity in recent times \cite {ref1,ref2}. Depending upon the strength of coupling and the nature of coupling, such systems are capable of entering into a state of phase (antiphase) \cite{ref3}, lag \cite{ref4}, anticipatory \cite{ref5}, generalized \cite{ref6} and complete synchronization \cite{ref7}. Among these, antiphase synchronization with repulsive or inhibitory coupling is the least studied. In the case of Rossler oscillators with normal diffusive coupling antiphase synchronization is reported for the funnel type attractors \cite{ref8}. So also cases of inhibitory (repulsive) coupling in coupled map lattices leading to synchronous, traveling wave and spatiotemporally chaotic states have been studied from the point of view of persistence \cite{ref9}. The role of repulsive coupling in enhancing synchronization in complex heterogenous networks has been reported recently\cite{ref10}. An array of coupled phase oscillators with repulsive coupling is found to result in a family of synchronized regimes with zero mean field \cite{ref11}.

The work reported in this paper is motivated by the fact that biological systems utilize different types of connections among them to realize synchronization or transmission of performances. For them the inhibitory coupling is as relevant and useful as excitatory coupling \cite{ref12} and biological networks often evolve with positive and negative connection between their components \cite{ref13}. An interesting case of agent-environment interaction using the dynamical systems theory has been explained, wherein the mutual interaction decides the adaptive fit of the agent \cite{ref14}. Our study is focused on the coupling between the systems and the environment and its effect in inducing synchronized behavior. Such an indirect coupling through environment has been studied in a system of van der Pol oscillators in the periodic state where negative feedback with environment results in complete synchronization while positive feedback leads to antiphase synchronization. The stability of the former state is analyzed in detail and applied to a model of pulsating secretion of GnRH \cite{ref15}. In the present work, we take Rossler oscillators in the chaotic state and show that antiphase synchronization occurs when the feedback from system to environment is negative (positive) and that from the environment to system is positive (negative). The relevance of the work lies in the fact that the systems are not in direct interaction with one another but get feedback from a dynamical environment which is influencing all systems. It is interesting to note that even this can induce a collective behavior which is phase correlated and can lead to zero mean field if the number of systems is large. Such a behavior can be found in many biological systems that communicate through their environment and often a large number of them synchronise to produce macroscopic oscillations of any desirable nature.

Our basic model system is n Rossler systems coupled indirectly through an environment with a dynamics governed by the following set of equations:
\begin{equation}
	\dot{x_k} = -y_k -z_k + \epsilon_1 w
\end{equation}
\begin{equation}
	\dot{y_k}= x_k + \alpha y_k
\end{equation}
\begin{equation}
	\dot{z_k}= \beta +z_k(x_k - \mu)
\end{equation}
\begin{math} 
        %\centering                             
	                    where (k=1,2,... n)               
\end{math}
\begin{equation}
	\label{1}
	\dot{w}=-w + \frac{\epsilon_2}{n}\sum (x_k) 
\end{equation}

where $(x_k, y_k, z_k)$ are the states of the oscillator, $(\alpha, \beta, \mu)$ are the oscillator parameters. Depending on parameter values the oscillators are in periodic or chaotic mode of operation. $w$ denotes the state of the environment, $\epsilon_1$, the coupling coeffient of feedback to the system and $\epsilon_2$ that of feedback to the environment. For simplicity we consider the feedback to the environment and vice versa through one of the variables $x_k$ only. The intrinsic dynamics of the environment is assumed to decay exponentially and without feedback from the oscillators it is incapable of sustaining itself for extended periods of time. We consider cases when $\epsilon_1 > 0$ and $\epsilon_2 < 0$ or $\epsilon_1 < 0$ and $\epsilon_2 > 0$, since when both $\epsilon_1, \epsilon_2 <0 $ or $\epsilon_1, \epsilon_2 >0 $ yields uninteresting or unbounded behavior.
 
For the numerical analysis we have taken $\alpha = \beta = 0.1$ and $\mu=18$ where the individual systems are chaotic. The system is evolved using fourth order Range Kutta Method for a time of 20,000 secs. As displayed in Fig 1(a) the states $x_1(t)$ and $x_2(t)$ are found to be 180 degrees out of phase. The similarity function \cite{ref16} $S^2(\tau)$, defined as 

\begin{equation}
S^2(\tau) = \frac{<[x_2(t+\tau)-x_1(t)]^2>}{[<x_1^2 (t)><x_2^2 (t)>]^1/2}
\end{equation}

is computed for $\tau=(0,21)$. The $x_1(t), x2(t)$ and $S^2(\tau)$ plot for two Rossler oscillators are shown in Fig 1

\begin{figure}
%\centering
\includegraphics[width=0.8\columnwidth]{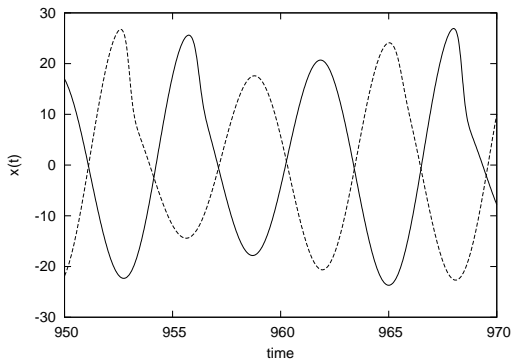}
\caption{ Time series plot of $x_1(t)$ and $x_2(t)$} 
\label{f.1.a}
\end{figure}
\begin{figure}
\includegraphics[width=0.8\columnwidth]{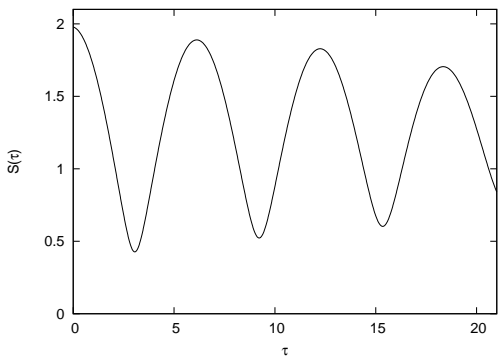}
\caption{Similarity function plot}
\label{f.1.b}
\end{figure}

It is clear that the first minima of $S^2(\tau)$ corresponds to a $\tau$ value which is the shift between $x_1(t)$ and $x_2(t)$. This minimum repeats with an average periodicity of nearly 6.0 seconds that corresponds to the approximate periodicity of the two individual oscillators. 

The above behavior is observed for the range $\epsilon_1=(0.01,0.95)$ and $\epsilon_2=(-0.5,-1)$ . Moreover when $\mu=4$ when the individual systems are in periodic state, qualitatively similar behavior with periodically synchronized states are found to occur. In this case the minimum of $S^2(\tau)$ is exactly zero and has a periodicity of 6.0 equal to the periodicity of the two oscillators.
\begin{figure}
%\centering
\includegraphics[width=0.8\columnwidth]{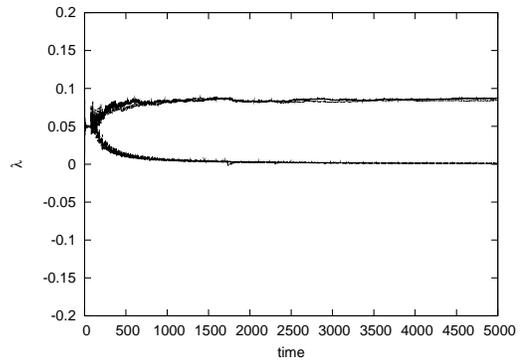}
\caption{The first four Lyapunov exponents of two independent Rossler systems for $\epsilon_1=0$ and $\epsilon_2=0$ in equation(1)}
\label{f.2}
\end{figure}

\begin{figure}
%\centering
\includegraphics[width=0.8\columnwidth]{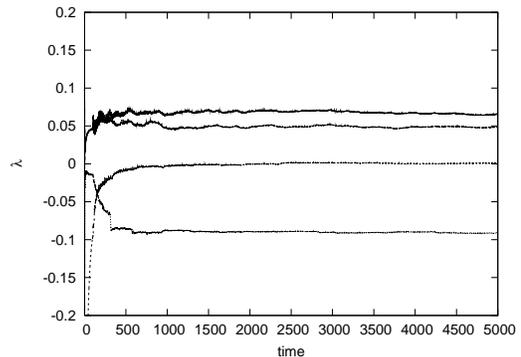}
\caption{The first four Lyapunov exponents of two environmentally coupled Rossler systems for $\epsilon_1=0.5$ and $\epsilon_2=-0.5$}
\label{f.3}
\end{figure}
The analysis is continued by calculating the Lyapunov spectrum for the 7-dimensional system given by equation (1) for n=2. When $\epsilon_1 = \epsilon_2 = 0$, the asymptotic values of the Lyapunov Exponents are $\lambda_1=0.084, \lambda_2=0.084, \lambda_3=0.001, \lambda_4=0.001, \lambda_5=-1.000, \lambda_6=-17.880$ and $\lambda_7=17.889$. The $\lambda_1, \lambda_2$ are positive with values corresponding to the $\lambda_1$ of an independent Rossler system \cite{ref17}. $\lambda_3$ and $\lambda_4 \approx 0$ and $\lambda_5, \lambda_6$ and $\lambda_7$ are negative. When $\epsilon_1=0.5 and \epsilon_2=-0.5$, corresponding to the antiphase synchronised case, we find that  $\lambda_1=0.064, \lambda_2=0.015, \lambda_3=0.000 \lambda_4=-0.075, \lambda_5=-0.089, \lambda_6=-10.927$ and $\lambda_7=-24.768$. Here one of the zero Lyapunov exponents becomes negative due to coupling, indicating phase correlation between the sub-systems \cite{ref1}. $\lambda_1$ and $\lambda_2$ though different still remain positive indicating that the amplitudes of the systems are uncorrelated and chaotic. The behavior of the first four Lyapunov exponents in both cases are given in Fig 2 and Fig 3.

Extending the analysis to higher number of systems coupled in the same way, we find that their phases are distributed over $2\pi$ with a lag of $2\pi /n$ between neighbors. This would mean that for large n, the mean field is zero. Similar results has been reported in \cite{ref11} in the context of directly coupled oscillators with repulsive coupling.The phase plots in x-y plane of the oscillator with a snapshot distribution of the systems marked as black dots is given in Fig 4 and Fig 5 for n=3 and 4. The similarity function and the spectrum of Lyapunov Exponents for n=3 and 4 are computed and are found to give qualitatively similar behavior as seen for n=2. We conclude that the onset of antiphase synchronization occurs when (n-1) zero Lyapunov exponents cross over to the negative region.
  
\begin{figure}
%\centering
\includegraphics[width=0.8\columnwidth]{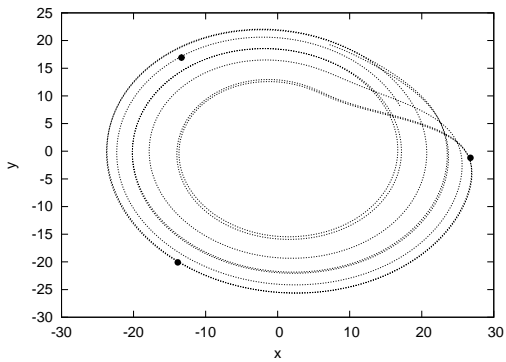}%
\caption{ The snapshot distribution of 3 Rossler Systems marked as black dots plotted on the x-y projection of the trajectory of a particular system(shown in dotted lines)}
\label{f.4}
\end{figure}

\begin{figure}
%\centering
\includegraphics[width=0.8\columnwidth]{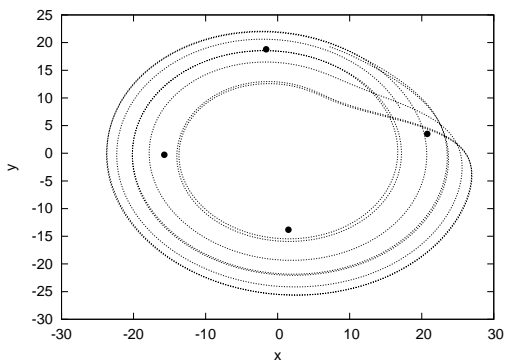}%
\caption{The snapshot distribution of 4 Rossler Systems marked as black dots plotted on the x-y projection of the trajectory of a particular system(shown in dotted lines)}
\label{f.5}
\end{figure}

In summary, we have studied the antiphase synchronization in a system of n Rossler oscillators that are coupled indirectly through a dynamic environment. The phase relation between the systems is analyzed via the time series and correlating it with the similarity function. The onset of phase correlated synchronized state is evident from the study of Lyapunov exponents also. When n is large, each of the systems are shifted in phase from its nearest neighbor by $2\pi/n$ resulting in zero instantenous mean field. By varying the coupling strength the same antiphase synchronization is observed with control of chaos. Further work aimed at synchronization of systems like neurons with environmental coupling is in progress and will be reported elsewhere.

%\env{acknowledgments}
{\bf {Acknowledgement}}\\
One of the authors, SV, thanks IISER, Pune for the facilities and warm hospitability provided during her summer project.

\end{document}